\documentclass[lettersize,journal]{IEEEtran}
\usepackage{amsmath,amsfonts}
\usepackage{algorithmic}
\usepackage{algorithm}
\usepackage{array}
\usepackage[caption=false,font=normalsize,labelfont=sf,textfont=sf]{subfig}
\usepackage{textcomp}
\usepackage{stfloats}
\usepackage{url}
\usepackage{verbatim}
\usepackage{pgfplots}
\usepackage{graphicx}
\usepackage{multirow}
\usepackage{booktabs}
\usepackage{tikz}
\usepackage{cite}
\usepackage{color,soul}
\begin{document}

\title{A Semi-Supervised Framework for Speech Confidence Detection using Whisper}

\author{\IEEEauthorblockN{Adam Wynn\IEEEauthorrefmark{1} and
Jingyun Wang\IEEEauthorrefmark{1}}

\IEEEauthorblockA{\IEEEauthorrefmark{1}Department of Computer Science,
Durham, United Kingdom, DH1 3DF}

\thanks{This paper was produced by the IEEE Publication Technology Group. They are in Piscataway, NJ.}
\thanks{Manuscript received April 19, 2021; revised August 16, 2021.}}

\markboth{Journal of \LaTeX\ Class Files,~Vol.~14, No.~8, August~2021}%
{Shell \MakeLowercase{\textit{et al.}}: A Sample Article Using IEEEtran.cls for IEEE Journals}

\IEEEpubid{0000--0000/00\$00.00~\copyright~2021 IEEE}

\maketitle

\begin{abstract}
Automatic detection of speaker confidence is critical for adaptive computing but remains constrained by limited labelled data and the subjectivity of paralinguistic annotations. This paper proposes a semi-supervised hybrid framework that fuses deep semantic embeddings from the Whisper encoder with an interpretable acoustic feature vector composed of eGeMAPS descriptors and auxiliary probability estimates of vocal stress and disfluency. To mitigate reliance on scarce ground truth data, we introduce an Uncertainty-Aware Pseudo-Labelling strategy where a model generates labels for unlabelled data, retaining only high-quality samples for training. Experimental results demonstrate that the proposed approach achieves a Macro-F1 score of 0.751, outperforming self-supervised baselines, including WavLM, HuBERT, and Wav2Vec 2.0. The hybrid architecture also surpasses the unimodal Whisper baseline, yielding a 3\% improvement in the minority class, confirming that explicit prosodic and auxiliary features provide necessary corrective signals which are otherwise lost in deep semantic representations. Ablation studies further show that a curated set of high confidence pseudo-labels outperforms indiscriminate large scale augmentation, confirming that data quality outweighs quantity for perceived confidence detection.
\end{abstract}


This work has been submitted to the IEEE for
possible publication. Copyright may be transferred
without notice, after which this version may no
longer be accessible.

\section{Introduction}
\IEEEPARstart{E}{ffective} communication is fundamental to human social interaction, enabling us to share knowledge, emotion, and intent \cite{Krauss1996}. Central to these interactions, speaker confidence serves as an important affective cue that directly influences listener perceptions of credibility, persuasiveness, and competence \cite{Mardiana2024,Guyer2019}. Therefore, the automatic modelling of confidence presents significant opportunities within the field of affective computing. For example, agents capable of detecting confidence can provide adaptive scaffolding to support self-efficacy in learning environments \cite{Cavalcanti2021}. Furthermore, the implications of confidence detection extend to mental health monitoring and empathetic Human-Computer Interaction (HCI), as low confidence is a known precursor to anxiety and social withdrawal \cite{Clark1995}.

Detecting confidence is a challenging task as it involves analysing both the semantic content and vocal delivery. A primary difficulty lies in resolving verbal and non-verbal discrepancies, situations where the text and audio features conflict. For example, a speaker may use assertive words, but their prosody indicates uncertainty through hesitation or pitch instability. Accurately identifying these discrepancies is essential for reliable confidence detection. Prior research has established that acoustic features such as pitch instability (jitter), intensity, dynamics, and speech rate vary with confidence levels \cite{Jiang2014}. However, automating this detection remains an open challenge. Most existing approaches have relied on small, manually annotated datasets and handcrafted features \cite{Pon2011}, which limits scalability. Furthermore, modern end-to-end approaches using deep speech representations (e.g., Wav2Vec 2.0, HuBERT) often prioritise lexical or phonemic fidelity over subtle prosodic dynamics \cite{Nair2020}, potentially failing to capture the subtle nuances of features that correlate with a lack of confidence.

This limitation is exacerbated by the scarcity of labelled data. Unlike basic discrete emotions, such as happiness, sadness and anger, which benefit from large-scale benchmarks, confidence datasets are small, subjective, and difficult to access. To address this, in this paper, a refined speech confidence dataset is constructed by re-annotating subsets of existing corpora including TED-LIUM \cite{Tedlium}, SEP-28K \cite{SEP28K}, CMU-MOSI\cite{CMUMOSI} and MLCommons People's Speech\cite{PeoplesSpeech}. This approach ensures coverage across diverse speaking styles and demographics whilst providing high-quality ground truth labels. However, since large-scale manual annotation is time-consuming, and to mitigate the reliance on a limited amount of annotated examples, a semi-supervised pseudo-labelling framework is employed. By leveraging a model to automatically label a larger unlabelled corpus, the training distribution is significantly expanded without incurring the cost of additional manual labelling.

\IEEEpubidadjcol
This paper extends our preliminary work \cite{WynnAIED2025} regarding semi-supervised confidence detection, where the integration of neural embeddings from the Whisper encoder \cite{Whisper} with a set of handcrafted features was previously explored. Whisper was prioritised over acoustic self-supervised learning models such as Wav2Vec 2.0 \cite{wav2vec2.0} due to its massive weakly-supervised pre-training, which yields representations that are semantically richer and more robust to speaker variability. Whilst the initial framework demonstrated the potential of pseudo-labelling \cite{psuedolabel} to mitigate data scarcity, this paper more rigorously evaluates the necessity of both the architecture and the learning strategy, and introduces a more diverse and expanded ground-truth dataset with increased annotator reliability. Moreover, this work incorporates an expanded feature vector consisting of eGeMAPS functionals \cite{eGeMaps} along with disfluency and stress auxiliary models to generate more robust pseudo-labels, and employs a Late Fusion strategy to integrate semantic and acoustic features without compromising the robustness of the semantic representation.

Therefore, a robust hybrid framework is proposed which fuses deep semantic embeddings from Whisper with a set of interpretable acoustic descriptors. Furthermore, to address data scarcity, an uncertainty-aware pseudo-labelling strategy that expands the training set by filtering for samples with the highest model certainty is introduced. This approach ensures that only high-quality samples are seen by the model, demonstrating that a small, curated curriculum of augmented data is significantly more effective than indiscriminate large-scale supervision.

Through this work, we address the following research questions:

\textbf{RQ1:} To what extent does the augmentation of training data with pseudo-labels yield a significant improvement in performance compared to a baseline trained exclusively on ground truth data?

\textbf{RQ2:} To what extent does prioritising data quality via uncertainty-aware filtering outweigh data quantity compared to standard pseudo-labelling in mitigating class imbalance and improving model robustness?

\textbf{RQ3:} To what extent does fusing explicit paralinguistic cues with deep semantic embeddings (Whisper) improve the detection of uncertainty compared to Whisper-only baselines?

In this paper, we advance the modelling of confidence through the following contributions. First, a hybrid framework is introduced which combines the semantic capabilities of Whisper embeddings with interpretable, handcrafted prosodic features. By employing a late-fusion strategy, this framework captures nuances of uncertainty often missed by deep encoders alone. Second, an Uncertainty-Aware Data Strategy is introduced to prioritise label quality over quantity. By filtering pseudo-labels based on labeller confidence, it is shown that the addition of a curated set of high-quality samples significantly improves generalisation compared to noisy large-scale augmentation. This research contributes to the broader field of speech processing by establishing one of the first dedicated frameworks for the automatic detection of speaker confidence.

\section{Related Work}

\subsection{Confidence and Disfluency Detection}
Confidence in speech is a complex paralinguistic construct perceived through a combination of vocal characteristics including pitch, dynamics, intensity, and articulation rate. Specifically, in this paper, confidence is defined as perceived speaker confidence rather than the speaker's internal state of self-efficacy or physiological stress. From a theoretical perspective, confidence maps closely to the Dominance dimension of the Valence-Arousal-Dominance (VAD) model \cite{Mehrabian1996}. Whilst many basic emotions (e.g., happiness or sadness) are distinguished primarily by valence and arousal \cite{Russell1997}, confidence is fundamentally characterised by high Dominance and Control \cite{Chappuis2022}. This creates a significant ambiguity in the signal space as acoustic markers of dominance frequently overlap with those of emotions such as anger, making it difficult to distinguish confidence from other high-arousal states based on prosody alone.

To identify these specific signals, Jiang and Pell \cite{Jiang2014, JIANG2017106 } mapped the primary vocal characteristics of confidence. Their research demonstrated that confident speech is characterised by a higher $f_0$ (pitch) range and mean amplitude, whilst unconfident speech exhibits a reduced speaking rate, lower mean $f_0$, and frequent pauses.

Despite this theoretical grounding, the transition to automated detection has been gradual. Early systems focused primarily on general public speaking skills rather than the specific affective state of confidence. For instance, Trinh et al. \cite{Trinh2017} developed RoboCop, an automated coaching system that provides feedback on speech quality metrics such as pacing and volume. While effective for general performance scoring, such systems often rely on heuristic rules rather than learning the latent representations of confidence.

More recent approaches have adopted deep learning to model perceived confidence. Chanda et al. \cite{Chanda2021} proposed a deep audiovisual framework using Bi-Directional LSTMs (Bi-LSTM) to classify confidence into three levels (High, Medium, Low) from interview recordings. Their work demonstrated that fusing audio and visual modalities yields measurable performance gains over unimodal baselines. However, their reliance on a small, specific dataset (34 candidates) and older recurrent architectures limits the generalisability of their findings to in-the-wild scenarios. Similarly, Nair et al. \cite{Nair2020} achieved 86.3\% accuracy using a CNN trained on MFCCs, but the reliance on a small, private dataset further restricts scalability. Our work addresses these gaps by employing a semi-supervised framework anchored by the Whisper encoder which improves generalisability compared to previous methods, enabling robust confidence detection across different speakers despite the scarcity of labelled data.

Confidence also heavily influences speech fluency. A lack of confidence is strongly correlated with an increase in disfluencies, such as interjections and prolongations \cite{Astuti2024}. Consequently, disfluency detection serves as a critical auxiliary task.  Current approaches to detecting speech disfluency heavily rely on annotated data, which is limited in availability. Among the existing datasets are SEP-28K \cite{SEP28K}, collected from online podcasts, and Fluencybank \cite{FluencyBank}, a dataset for the study of fluency development of native and non-native adults and children.
Several approaches for disfluency classification have emerged including  passing spectrograms into residual networks followed by a bi-directional LSTM for stutter classification \cite{Kourkounakis20}, or using Transcription Based Methods \cite{Boughariou2024}. Still, these approaches require large labelled datasets.

To address the challenge of limited data, Mohapatra et al. \cite{Mohapatra2022} proposed DisfluencyNet, a Wav2Vec 2.0-based model with convolutional and fully connected layers, achieving over 5\% improvement in disfluency detection compared to baselines with only a few minutes of data. Similarly, Liu et al. \cite{Liu2023Disfluency} added a Transformer layer to enhance Wav2Vec 2.0's generalisation for detecting disfluencies across languages. Ameer et al. \cite{WhisperDisfluency} used the Whisper model for multi-class disfluency classification, introducing an encoder-freezing strategy and refining the SEP-28K dataset. They demonstrated that Whisper outperformed Wav2Vec 2.0 in multi-class disfluency tasks, achieving an average F1 of 0.81, offering better generalisation and faster inference. However, their approach struggled to accurately classify fluent speech (F1=0.23) suggesting that multi-class labels introduce significant noise. Therefore, in this paper, disfluency detection is treated as a binary-classification task using the Whisper encoder, providing a useful auxiliary signal for detecting reduced speaker confidence.

\subsection{Speech Feature Extraction and Speech Emotion Recognition}
Confidence detection shares significant methodological overlap with Speech Emotion Recognition (SER) as both tasks rely on disentangling prosodic affect from linguistic content. Traditional SER approaches use handcrafted features such as Mel Frequency Cepstral Coefficients (MFCCs) \cite{tiwari2010mfcc}, which capture the spectral envelope but often fail to model long-term temporal dynamics \cite{Sidhu2024-ed}.

To capture these missing details, modern systems \cite{Chen2021} have adopted the eGeMAPS \cite{eGeMaps} feature set. Unlike MFCCs, which only describe the general shape of the sound, eGeMAPS encodes specific, meaningful vocal patterns such as voice shakiness (jitter) and volume changes which are close to what listeners rely on to gauge a speaker's perceived confidence. Despite their interpretability, these features often struggle to generalise across diverse acoustic environments or capture the high-level semantic context of an utterance.

To address these limitations, the introduction of Self-Supervised Learning allows models to automatically extract rich, high-dimensional representations from audio. Pepino et al. \cite{Pepino2021} demonstrated that Wav2Vec 2.0 embeddings, when fused with prosodic features, outperform handcrafted baselines in SER. However, Wagner et al. \cite{wagner2022} caution that transformer-based models can become over-reliant on linguistic information, potentially ignoring paralinguistic cues when the text is semantically dominant. This semantic bias is particularly relevant for confidence detection, where a speaker may say something with high semantic confidence but with an unconfident-sounding voice. To address this, recent studies have explored the weakly-supervised Whisper model for paralinguistic tasks. Goron et al. \cite{Goron2024} and Osman et al. \cite{osman2024} found that Whisper embeddings generalise better to unseen speakers than Wav2Vec 2.0 or HuBERT, likely due to the massive diversity of its weak-supervision training data.

However, because ASR models like Whisper are optimised to minimise Word Error Rate, they naturally prioritise linguistic invariance over prosodic variability. Consequently, fine-grained cues like pitch instability may be treated as secondary to the lexical content. Our hybrid approach complements the strong semantic backbone of Whisper by explicitly reintroducing these paralinguistic signals by fusing explicit paralinguistic cues with the Whisper encoder embeddings.

\subsection{Semi-Supervised Learning in Speech Processing}

The scarcity of labelled data necessitates data-efficient learning strategies. Whilst Transfer Learning from large pre-trained models (e.g., Whisper, Wav2Vec 2.0) provides strong initial representations, adapting these models to specific downstream tasks typically requires fine-tuning on labelled data, which remains a bottleneck. To address this, semi-supervised learning techniques such as Pseudo-Labelling \cite{psuedolabel} leverage a model to generate target labels for unlabelled data, expanding the overall amount of available data. However, standard pseudo-labelling is prone to confirmation bias, where the hybrid model overfits to the labellers' incorrect predictions \cite{arazo2020pseudo}. This risk is exacerbated in affective computing, where ground-truth labels are subjective and class boundaries are ambiguous. 

To mitigate this, recent literature emphasises uncertainty-aware filtering. Approaches such as FixMatch \cite{sohn2020fixmatch} use strict confidence thresholds to ensure that only high-quality samples contribute to the model. In the speech domain, this approach is often extended via Cross-View Supervision, where models trained on different 'views' of the data supervise each other. We adopt this philosophy by using a separate pseudo-labeller model trained only on acoustic features such as eGeMAPS, rather than allowing the Whisper model to teach itself. This prevents the system from reinforcing its own semantic biases. By generating labels based on explicit vocal cues, we create a high-precision dataset that is robust to the noise often found in real-world audio.

\section{Methodology}

\begin{figure*}[t!]
\centering
\includegraphics[width=17cm]{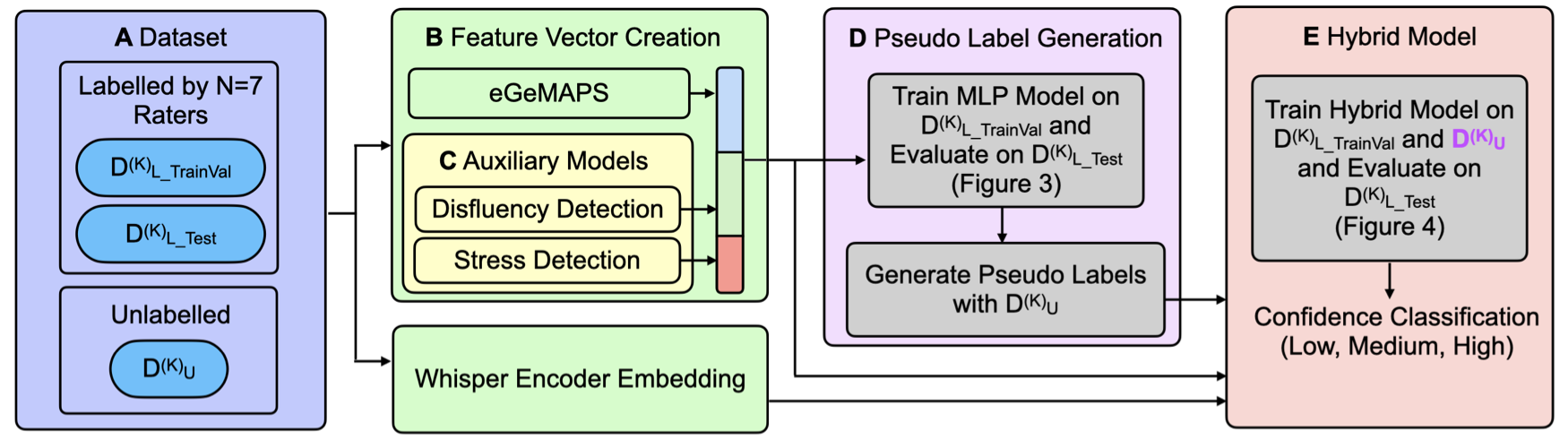}
\caption{Overall Pipeline of the Confidence Classification System during Training for Fold k} \label{fig:pipeline}
\vspace{-5mm}
\end{figure*}

Motivated by recent progress in deep learning approaches to audio representation and classification, this study proposes a hybrid framework for speech confidence detection. As illustrated in Fig. \ref{fig:pipeline}, the method integrates interpretable prosodic and spectral features with deep audio embeddings derived from the Whisper-base encoder.  The entire training and evaluation protocol is executed via a robust 5-Fold Cross-Validation Pipeline which comprises five key stages: (A) Dataset Creation and Annotation, (B) Feature Vector Creation, (C) Auxiliary Disfluency and Stress Detection, (D) Model-based pseudo-label generation, and (E) Hybrid model training for confidence detection which incorporates information from both the feature vector and deep audio embeddings effectively to develop a robust system for classifying speaker confidence levels on limited annotated data. 

The pipeline begins with the creation of a small annotated dataset labelled by human raters. To address the challenge of limited available labelled data, a pseudo-labelling approach is used to expand the dataset sufficiently to train the model effectively.

\subsection{Dataset and Audio Preprocessing}

Due to the absence of publicly available datasets and standard benchmarks for perceived confidence, a manual confidence annotation process was conducted. A custom dataset ($D_{L}$) comprising $N=600$ utterances between 5 and 12 seconds was curated by sampling from the TED-LIUM~\cite{Tedlium}, CMU-MOSI \cite{CMUMOSI},  MLCommons People’s Speech \cite{PeoplesSpeech}, and SEP-28K~\cite{SEP28K} datasets, supplemented with additional recordings of both native and non-native English speakers in beginner speech contests to ensure diversity. 

\begin{table}[h]
\caption{Inter-Rater Reliability Analysis ($N=576$, $k=7$ raters)}

\label{tab:icc}
\centering
\setlength{\tabcolsep}{2.5pt}
\begin{tabular}{l c c c c c c}
\hline

\textbf{Measure} & \textbf{ICC} & \textbf{95\% CI} & \textbf{F} & \textbf{df1} & \textbf{df2} & \textbf{p-value} \\
\hline
Single Rater & $0.48$ & $[0.44, 0.53]$ & $8.21$ & $575$ & $3450$ & $< 0.001$ \\
\textbf{Average (Consensus)} & $\mathbf{0.87}$ & $[0.85, 0.89]$ & $8.21$ & $575$ & $3450$ & $\mathbf{< 0.001}$ \\
\hline
\multicolumn{7}{l}{\footnotesize \textit{Model: Two-way random effects, absolute agreement (ICC 2,k).}}
\end{tabular}
\end{table}

As confidence in speech is  subjective, to ensure robust labels, each clip was independently annotated by 7 English-fluent annotators (4 male, 3 female) using a three-level ordinal scale: \emph{low}, \emph{medium}, and \emph{high}.  
We computed the Intraclass Correlation Coefficient (ICC) using a two-way random-effects model. Analysis was performed on $N=576$ samples. The remaining 24 samples were excluded from this specific calculation because one or more raters flagged the audio as 'Not Clear', resulting in incomplete ordinal data for those instances. However, these samples were retained in the final dataset ($N=600$) as a clear consensus label could still be derived from the remaining valid annotations.
Whilst single-rater agreement was moderate ($ICC=0.48$), reflecting individual subjectivity, the average-measures reliability was excellent ($ICC=0.87$), confirming that the collective consensus provides a stable training signal. Individual labels were aggregated using the Dawid-Skene (DS) probabilistic model~\cite{DawidSkene}, which further refines the ground truth by weighting annotators based on their consistency. The final curated dataset comprises 300 high, 210 medium, and 90 low-confidence clips.

\begin{figure}
\centering
\includegraphics[width=7cm]{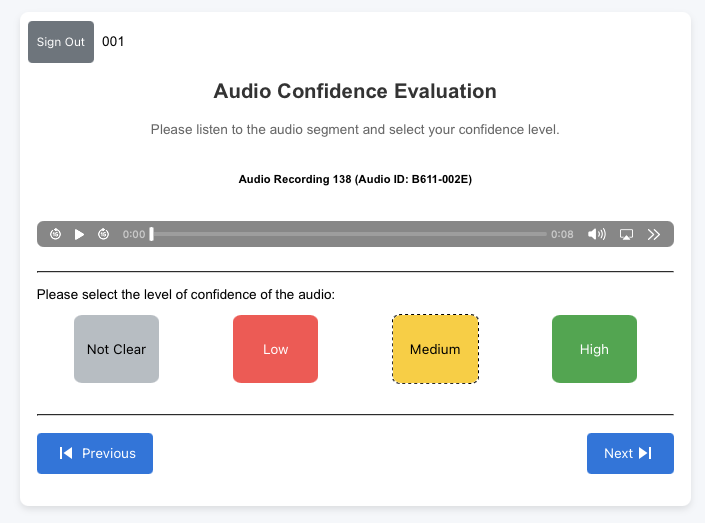}
\caption{User Interface of the Labelling System} \label{fig:labellingsystem}
\vspace{-5mm}
\end{figure}

To overcome the limitation of the dataset’s small size and inherent class imbalance, a model-based pseudo-labelling approach is applied to expand it. The augmented dataset, sampled from the same corpora as $D_L$, with all segments appearing in $D_L$ excluded, is then used for both training and validation (development). Further details on this technique are provided in Section \ref{sec:pseudolabel}. All audio files are converted to WAV format, resampled to 16 kHz, and converted to mono. Background noise is removed from each clip using the noisereduce Python library \cite{nrtim_sainburg_2019_3243139}.

As previously mentioned, to ensure rigorous evaluation without data leakage, the entire pipeline operates under a consistent stratified 5-fold cross-validation framework. The dataset $D_{L}$ is partitioned into 5 folds once and these splits remain fixed throughout the experiment. The training folds used to build the pseudo-label model ($D_{L\_TrainVal}^{(k)}$) are the exact same folds used to train the final hybrid model, and the held-out test sets ($D_{L\_Test}^{(k)}$) are identical for both evaluations. Importantly, for any given fold $k$, the held-out test set $D_{L\_Test}^{(k)}$ is never seen by either model during training.

\subsection{Feature Vector Creation}

For every input audio $x_i$, two distinct modalities are extracted.
(1) Whisper Base Encoder: The Whisper Base encoder is used to extract 512-dimensional semantic embeddings. The first three encoder layers are frozen to preserve linguistic representations.
(2) Feature Vector: 
We construct a 94-dimensional acoustic vector $\mathbf{f}_i$ containing eGeMAPS prosodic features (88 dim) and Auxiliary Scores (6 dim) for Disfluency and Stress.

\subsubsection{Acoustic and Prosodic Features (OpenSMILE)}
To capture  paralinguistic cues, we used the OpenSMILE toolkit \cite{Opensmile} to extract the eGeMAPSv02 (extended Geneva Minimalistic Acoustic Parameter Set) \cite{eGeMaps}. This standardised feature set consists of 88 functionals derived from  low-level descriptors across frequency, energy, spectral and temporal domains.

These features provide a robust, interpretable baseline for detecting perceived confidence, and are used as inputs for the pseudo-labelling models.

\subsubsection{Auxiliary High-Level Features}To augment the low-level acoustic descriptors, we append 6 high-level probability scores derived from auxiliary Whisper-based classifiers (Section III-C):$$\mathbf{f}_{aux} = [\mathbf{d}_i, s_i] \in \mathbb{R}^6$$where:\begin{itemize}\item $\mathbf{d}_i \in \mathbb{R}^5$ represents the calibrated probabilities for five distinct disfluency types: Block, Prolongation, Interjection, Word Repetition, and Sound Repetition.\item $s_i \in \mathbb{R}^1$ represents the calibrated probability of perceived vocal stress.\end{itemize}The final feature vector $\mathbf{f}_i = [\text{eGeMAPS}_i; \mathbf{d}_i; s_i]$ is z-score normalised using statistics from the training partition to ensure numerical stability during training.

\subsection{Auxiliary Disfluency and Stress Detection} 

\subsubsection{Disfluency Detection}
\label{sec:disfluency}
\begin{figure}

\centering
\includegraphics[width=4cm]{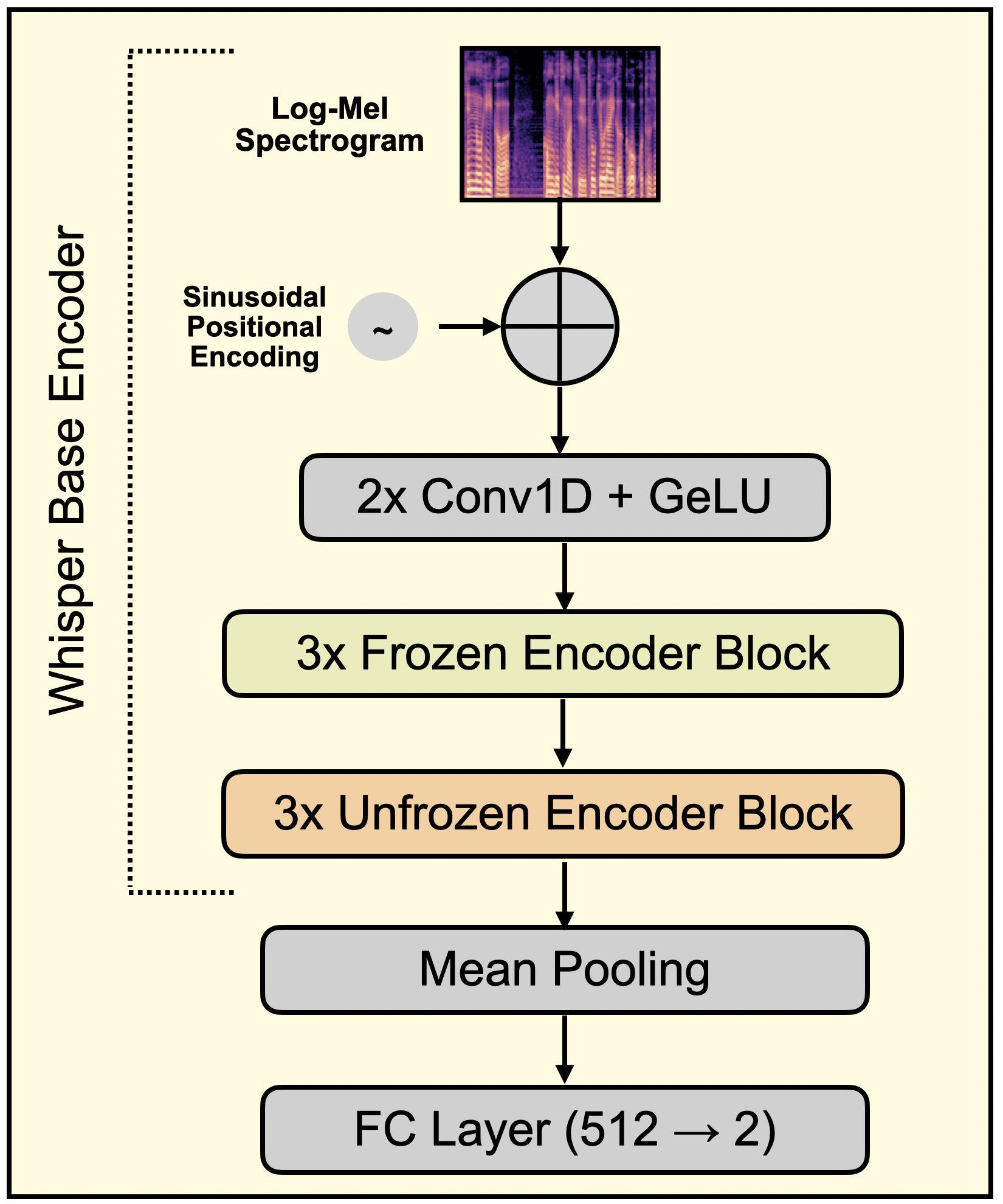}
\caption{Architecture of Auxiliary Models - Disfluency and Stress Classifiers} \label{fig:disfluency}
\end{figure}

Confidence can also influence the frequency of speech disfluencies, as speakers who feel less confident often exhibit more pauses, repetitions, and filler words \cite{Astuti2024}. Therefore, as disfluency patterns can serve as an indicator of confidence, detecting these disfluencies is an essential step in confidence detection.

\paragraph{Dataset}
Among existing English-language resources for disfluency detection, two primary datasets are widely used: SEP-28K \cite{SEP28K}, which consists of 28,000 three-second clips extracted from eight podcasts, and FluencyBank \cite{FluencyBank}, which contains approximately 4,000 instances designed for the study of fluency development in native and non-native adults and children. Each sample in these datasets was annotated for the presence of the disfluency types in table \ref{tab:disfluency_types}.

\begin{table}[h]
\centering
\caption{Disfluency types included in  SEP-28K and FluencyBank.}
\label{tab:disfluency_types}
\begin{tabular}{p{1.5cm}p{4cm}p{2cm}}
\toprule
\textbf{Type} & \textbf{Definition} & \textbf{Example} \\
\midrule
Prolongations & Elongation of speech sounds, often reflecting hesitation or planning. & ``ssssso'' \\[3pt]
Blocks & Temporary stoppage of airflow, producing audible silence before the next phoneme. & (pause) \\[3pt]
Sound Repetitions & Repetition of individual phonemes or syllables. & ``b-b-but'' \\[3pt]
Word Repetitions & Repetition of entire words, indicating hesitation or correction. & ``I I think'' \\[3pt]
Interjections & Filler or non-lexical utterances that signal hesitation or low confidence. & ``uh,'' ``um,'' ``like'' \\
\bottomrule
\end{tabular}
\end{table}


The SEP-28K-Extended (SEP-28K-E) variant \cite{SEP28KE} was introduced to mitigate speaker imbalance in the original dataset, which was dominated by four main podcast hosts. It separates training and evaluation speakers to improve cross-speaker generalisation. The SEP-28K-E-Merged version further integrates FluencyBank data for testing, enabling cross-corpus evaluation and improved robustness to differences in recording conditions and speaking styles. This merged dataset has therefore become a benchmark for disfluency detection due to its size, diversity, and well-defined splits. Therefore, in this study, we employ the SEP-28K-E-Merged dataset \cite{WhisperDisfluency}.

\paragraph{Data Preprocessing}
Each three-second audio clip in SEP-28K-E-Merged was labelled independently by three annotators so only the audio clips where the majority of raters agreed were used and data was excluded if any rater indicated they were unsure. Additionally, data was excluded which had poor audio quality, was difficult to understand, or contained music or non-speech content.

To examine the effect of class imbalance on disfluency detection, we train
models using two fluent:disfluent sampling ratios: 0.8 and 1.0. Whilst the training set was balanced to prevent bias toward the fluent majority class, the validation and test sets retain their natural class distributions to enable realistic performance evaluation

\paragraph{Model Architecture}

The model architecture (Fig. \ref{fig:disfluency}) uses HuggingFace's WhisperForAudioClassification implementation with the Whisper-base encoder \cite{Whisper} for binary disfluency detection. The Whisper-base encoder processes 80-dimensional log-mel spectrograms to produce contextualised acoustic representations. The first three encoder layers were frozen to preserve pre-trained linguistic representations. The classification head performs pooling over the encoder outputs and applies a linear transformation to produce two output logits for binary classification. Cross-entropy loss is used for optimisation during training with the AdamW optimiser \cite{AdamWOpt} (learning rate $2.5\times10^{-5}$;  weight decay $1\times10^{-5}$), and early stopping based on validation loss.

\subsubsection{Stress Detection}
\label{sec:stress}

\paragraph{Dataset}

 To train and evaluate the stress classification model, three publicly available emotional speech corpora were combined: RAVDESS \cite{RAVDESS}, SAVEE \cite{SAVEE}, and TESS \cite{TESS}. Each dataset contains recordings of actors expressing a range of emotions, which were re-labelled into low stress and high stress, to better align with the goals of confidence detection.
 
Following the affective mapping protocols established in prior work \cite{StressA, StressB}, in the RAVDESS dataset, emotions such as neutral, calm, and happy were labelled as low stress, while sad, angry, and surprised were labelled as high stress. The SAVEE dataset followed a similar mapping, with angry and sad labelled as high stress, and neutral and happy as low stress. In TESS, neutral and happy were considered low stress, and sad, angry, and surprised as high stress. After balancing, this results in 1460 low stress and 1460 high stress labels.

\paragraph{Data Preprocessing}

All audio samples were converted into 16 kHz, single-channel waveforms and normalised prior to Mel-spectrogram extraction. The combined dataset provided a diverse set of emotional speech instances across multiple speakers and recording conditions, enhancing the model’s robustness.

\paragraph{Model Architecture}

The stress classification model follows a similar architecture to the disfluency model, using HuggingFace's WhisperForAudioClassification implementation to distinguish between low and high stress levels from 80-dimensional log-mel spectrograms. 

However, due to the smaller number of samples per stress category compared to large disfluency corpora, the use of stratified 10-fold cross-validation with early stopping (patience of 10 epochs) was employed with 20\% data left out for testing, to ensure robust performance evaluation across the binary classification problem.

\subsubsection{Probability Calibration}

To ensure the reliability of these features for the downstream task  post-hoc temperature scaling~\cite{guo2017} was applied. This involved dividing the logits $\mathbf{z}$ by a scalar parameter $T$ before the softmax activation. This parameter $T$ is optimised to minimise negative log-likelihood on the validation set, ensuring the output probabilities accurately reflect the model's true confidence.

\subsection{Model-Based pseudo-Label Generation }
\label{sec:pseudolabel}

\begin{figure}

\centering
\includegraphics[width=4cm]{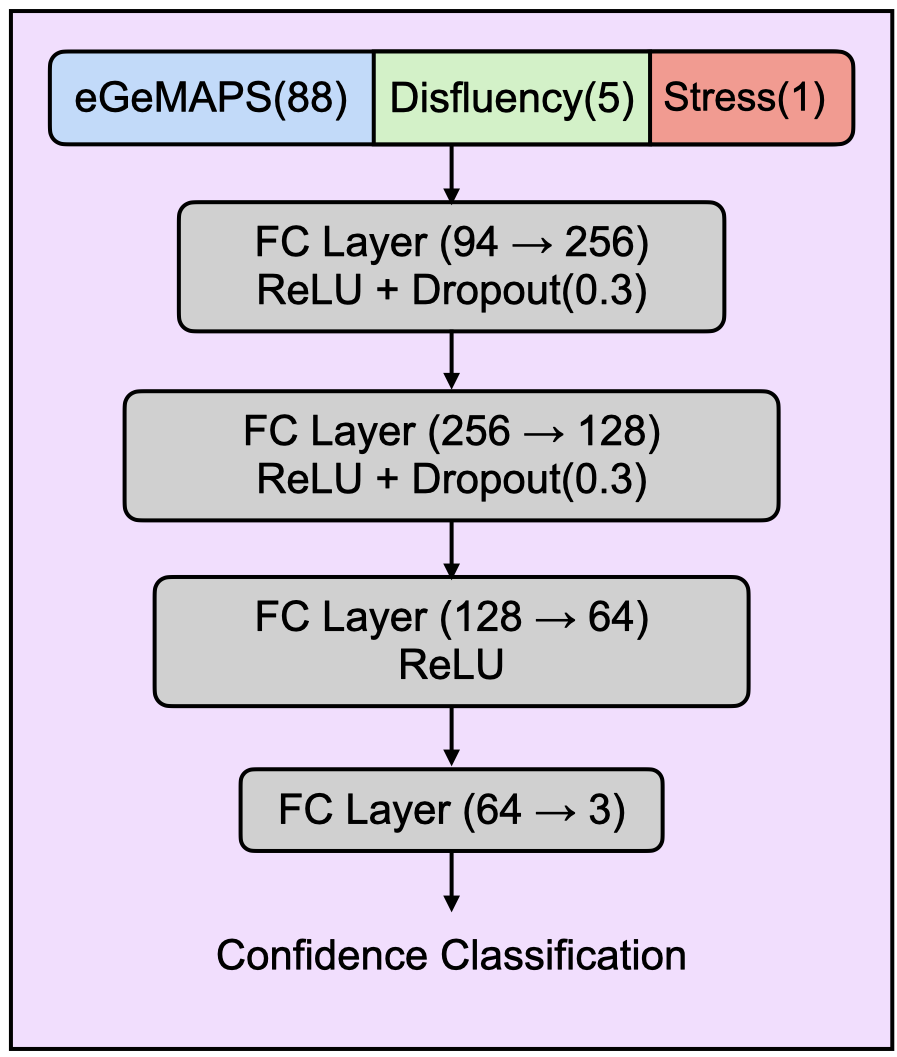}
\caption{Architecture of MLP Labeller} \label{fig:labeller}
\end{figure}

To address the limitation of a small ground truth dataset ($N_{GT}=600$), a model-based pseudo-labelling approach was used to generate additional labelled data by propagating labels from the ground truth to a larger unlabelled corpus. After the feature vectors  ($\mathbf{f}_{i}$) were created, a multi-layer perceptron (MLP) was trained on the ground truth dataset feature vectors to classify speech confidence into three levels: low, medium, and high, as shown in Figure \ref{fig:labeller}. The model was trained with Adam optimiser \cite{AdamOpt}, cross-entropy loss, and a learning rate of 0.001.
 This process did not involve the audio or the Whisper embeddings directly but instead relied only on the feature vector representations ($\mathbf{f}_{i}$) of each sample in an attempt to prevent data leakage. 

The trained MLP was then applied to the unlabelled corpus to generate probability distributions over the three confidence classes. To mitigate the risk of confirmation bias, where the main hybrid model learns the labeller's mistakes, we applied a strict confidence thresholding filter ($\tau$). Only samples where the model's prediction confidence exceeded the threshold were retained as pseudo-labels. Based on empirical ablation studies, we initially set $\tau = 0.8$. This filtering removes the ambiguous confidence samples where the MLP was uncertain, curating a high-precision dataset with on average $N\approx 1194 \pm 345$ audio samples across each fold from an initial pool of 10589 unlabelled segments.

 However, the filtering process naturally skews the distribution and low confidence samples were less frequent than High. To avoid physical downsampling in order to preserve the full diversity of the pseudo-labeled corpus, we used a Weighted Random Sampler that oversamples the minority classes ensuring that every training batch contains a uniform class distribution. The resulting pseudo-labelled dataset, combined with the original ground truth samples, provided a significantly larger dataset for subsequent stages of the model development pipeline, enabling improved generalisation and performance of the final confidence classification system.
\subsection{Hybrid Model Training}

\begin{figure}

\centering
\includegraphics[width=9cm]{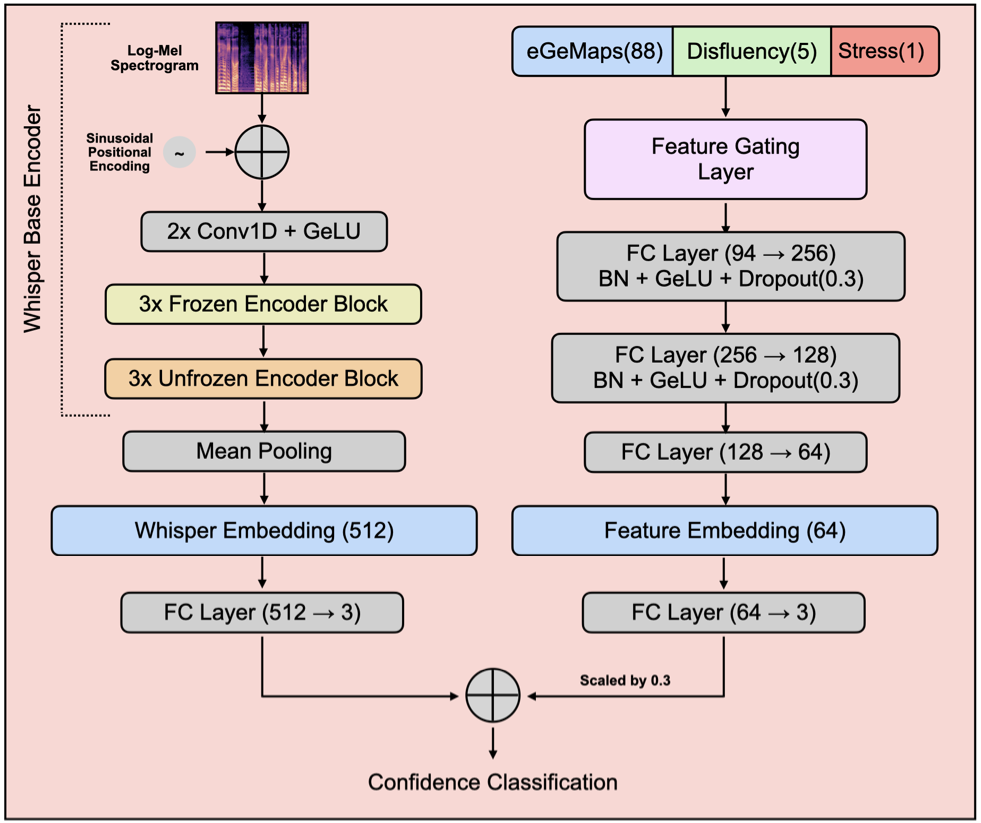}
\caption{Architecture of Hybrid Confidence Model} \label{fig:hybrid}
\end{figure}

The final stage of the pipeline involves training the hybrid model. As illustrated in Fig. \ref{fig:hybrid}, this model integrates the deep semantic capabilities of Whisper with the interpretable acoustic signals from the feature vector.

\subsubsection{Architecture}
The model operates via two parallel processing streams:
\begin{itemize}
    \item Whisper Stream: The pre-processed audio is input into the Huggingface implementation of the  Whisper-base encoder. The resulting 512-dimensional embeddings are passed through a linear projection head to generate semantic logits. To prevent overfitting and preserve pre-trained linguistic knowledge, the first three encoder layers are frozen.
    \item Feature Vector Stream: The 94-dimensional feature vector $\mathbf{f}_i$ is passed through a Feature Gating Layer which applies a learnable sigmoid mask to suppress irrelevant features  before the vector is further processed by the MLP. As shown in Figure \ref{fig:hybrid}, the first two fully connected layers of the MLP are each followed by Batch Normalisation and GELU activation to stabilise training dynamics, along with a Dropout layer ($p=0.3$) to prevent overfitting on the limited ground truth data.
\end{itemize}

To combine the two modalities, a late fusion strategy is used, where the final prediction is derived from a weighted sum of the logits. A scaling factor of $\lambda=0.3$ is assigned to the feature vector so that Whisper provides the base confidence estimate, whilst the Feature Vector acts as a corrective signal.

\subsubsection{Training Strategy}
The model is trained on the union of the fold-specific ground truth and the pseudo-labelled dataset ($D_{L\_TrainVal}^{(k)} \cup D_{U}^{(k)}$). To ensure the model prioritises verified human annotations despite the volume of pseudo-labels, we employ a Source-Boosted Loss Function:
\begin{equation}
\mathcal{L} = \omega_{class} \cdot (\mathcal{L}_{CE}(y_{L}, \hat{y}) \cdot 18.0 + \mathcal{L}_{CE}(y_{U}, \hat{y}))
\end{equation}
where the ground truth loss is scaled by a factor of 18. This scaling factor was determined empirically to normalise the gradient contribution of the two datasets. Additionally, a class weight $\omega_{med}=1.2$ ($\omega_{low,high}=1.0$ ) is applied to the minority Medium class to improve boundary separation. Optimisation is performed using AdamW \cite{AdamWOpt} with a cosine annealing scheduler. A differential learning rate strategy is employed where the pre-trained Whisper stream is fine-tuned with a lower rate of $2.5 \times 10^{-5}$ whilst the MLP feature stream is trained with a rate of $1 \times 10^{-3}$.

In our preliminary work \cite{WynnAIED2025}, we additionally employed a complex downstream sequence model. However, this study demonstrates that pre-trained semantic features are sufficient for direct mapping. This simplification reduces trainable parameters while improving F1 score, highlighting the efficiency of the proposed model.

\section{Results}

\subsection{MLP Labeller}
To generate pseudo-labels for the final hybrid model, a Multi-Layer Perceptron (MLP) was first trained on the engineered feature vector ($f_i$) only. For each outer fold k, the MLP was trained on the training partition of the ground truth dataset ($D^{(k)}_{L\_TrainVal}$) using an internal 5-fold cross-validation and its performance is reported on the held-out Test Set ($D^{(k)}_{L\_Test}$). 

On the test sets ($D^{(k)}_{L\_Test}$) the MLP labeller achieved a mean macro F1-score of 0.746. The MLP demonstrated stronger capability in distinguishing low (F1=78.2) and high confidence (F1=81.8) but struggled more with the middle confidence class (F1=64.4), where misclassifications were primarily concentrated between adjacent classes (e.g., Medium misclassified as High, rather than Low misclassified as High). Therefore, as mentioned in section \ref{sec:pseudolabel} a confidence threshold ($\tau$) is applied to filter out the ambiguous labels where the MLP was less confident in its predictions, creating a high-precision dataset to supervise the downstream hybrid model.

\subsection{Auxiliary Disfluency and Stress Detection Results}

To validate the quality of the auxiliary features fed into the hybrid model, the performance of the disfluency and stress detectors are evaluated in this section.

\subsubsection{Disfluency Detection Results}

To evaluate the effectiveness of different design choices in the disfluency detection component, an ablation study optimising model capacity (Whisper-Base vs. Tiny), fine-tuning strategy (Frozen vs. Unfrozen encoder), and class balancing was conducted. All studies were conducted on the SEP-28-K-E-Merged Test Set.

As shown in Table~\ref{tab:disfluency_ablation}, the best-performing result was using the Whisper-Base model with frozen encoder layers and a balance ratio of 0.8.
 Among the five disfluency categories, interjections achieved the highest performance (F1 = 0.90), followed by sound repetitions (F1 = 0.81) and prolongations (F1 = 0.73). In contrast, blocks and word repetitions were more difficult to classify, reaching F1-scores below 0.65. This pattern aligns with prior findings that such disfluencies are often acoustically subtle or context-dependent, making them harder to detect automatically~\cite{Mohapatra2022, Liu2023Disfluency}. Moreover, regarding class balance, the 0.8 ratio (mild imbalance) achieved the highest F1 (0.77), outperforming the strictly balanced (1.0) setup, suggesting that moderate balancing better preserves natural speech variability and improves robustness.

Across all disfluency types and balance ratios, as shown in table \ref{tab:agreement_full_f1}, the best performing model uses labels where 2+ raters agree yields higher F1-scores than when all 3 raters agree.  
Whilst some prior work on SEP-28K \cite{Liu2023Disfluency, Mohapatra2022} suggests that unanimous labels yield higher annotation quality, our results show that the reduction in dataset size associated with this harms model generalisation, particularly for low-frequency disfluencies such as blocks and prolongations. 

\begin{table*}[!ht]
\centering
\caption{F1-scores for each disfluency type under different model configurations and balance ratios.}
\label{tab:disfluency_ablation}
\begin{tabular}{lcccccccc}
\toprule
\multirow{2}{*}{\textbf{Type}} & 
\multicolumn{2}{c}{\textbf{Base Frozen}} &
\multicolumn{2}{c}{\textbf{Base Unfrozen}} &
\multicolumn{2}{c}{\textbf{Tiny Frozen}} &
\multicolumn{2}{c}{\textbf{Tiny Unfrozen}} \\ 
\cmidrule(lr){2-3}\cmidrule(lr){4-5}\cmidrule(lr){6-7}\cmidrule(lr){8-9}
 & 0.8 & 1.0 & 0.8 & 1.0 & 0.8 & 1.0 & 0.8 & 1.0 \\
\midrule
Blocks & 0.642 & \textbf{0.664} & 0.591 & 0.582 & 0.554 & 0.536 & 0.562 & 0.533 \\
Interjections & \textbf{0.900} & \textbf{0.900} & 0.892 & 0.884 & 0.883 & 0.881 & 0.873 & 0.862 \\
Prolongations & \textbf{0.730} & 0.685 & 0.657 & 0.690 & 0.672 & 0.638 & 0.617 & 0.652 \\
Sound Repetitions & \textbf{0.813} & 0.793 & 0.795 & 0.763 & 0.724 & 0.642 & 0.701 & 0.670 \\
Word Repetitions & 0.743 & \textbf{0.760} & 0.729 & 0.709 & 0.644 & 0.631 & 0.591 & 0.577 \\
\midrule
\textbf{Mean} & \textbf{0.766} & 0.760 & 0.733 & 0.726 & 0.695 & 0.666 & 0.669 & 0.659 \\
\bottomrule
\end{tabular}
\end{table*}

\begin{table}[t]
\centering
\caption{F1-scores for Whisper-Base disfluency detection under different annotation agreement conditions.}
\label{tab:agreement_full_f1}
\begin{tabular}{lcccc}
\toprule
& \multicolumn{2}{c}{\textbf{2+ Raters Agree}} & \multicolumn{2}{c}{\textbf{All 3 Raters Agree}} \\
\cmidrule(lr){2-3} \cmidrule(lr){4-5}
\textbf{Disfluency Type} & \multicolumn{2}{c}{Base Frozen} & \multicolumn{2}{c}{Base Frozen} \\
& 0.8 & 1.0 & 0.8 & 1.0 \\
\midrule
Block & 0.642 & \textbf{0.664} & 0.289 & 0.274 \\
Interjection & \textbf{0.900} & 0.900 & 0.823 & 0.830 \\
Prolongation & \textbf{0.730} & 0.685 & 0.643 & 0.681 \\
Sound Repetition & \textbf{0.813} & 0.793 & 0.664 & 0.734 \\
Word Repetition & 0.743 & \textbf{0.760} & 0.717 & 0.740 \\
\midrule
\textbf{Mean F1} & \textbf{0.766} & 0.760 & 0.627 & 0.652 \\
\bottomrule
\end{tabular}
\end{table}

\subsubsection{Stress Detection Results}

Table \ref{tab:stress_results} presents the mean 10-fold cross-validation performance of the four
Whisper-based stress classifiers using stratified 10-fold cross-validation. Results are reported strictly on the held-out test set. All configurations achieve similarly strong results, with F1-scores ranging from 0.936 to 0.942.
\begin{table}[t]
\centering
\caption{Stress detection performance across Whisper variants.}
\label{tab:stress_results}
\begin{tabular}{lcc}
\toprule
\textbf{Model} & \textbf{Frozen} & \textbf{F1-Score} \\
\midrule
Whisper Base  & Yes & 0.9385 \\
Whisper Base  & No  & \textbf{0.9423} \\
Whisper Tiny  & Yes & 0.9413 \\
Whisper Tiny  & No  & 0.9357 \\
\bottomrule
\end{tabular}
\end{table}

Given the negligible performance gap, we adopt Whisper-Base with frozen
lower layers for the remainder of the experiments to ensure
architectural consistency with the disfluency detector and reduce computational cost during deployment by avoiding unnecessary fine-tuning.

\subsection{Hybrid Model Results}

\subsubsection{Experimental Setup}

To ensure a robust evaluation of the proposed hybrid model, as explained in section III, a stratified 5-Fold Cross-Validation scheme was used. In each fold, the model was trained on a combination of the ground truth dataset ($D_{L\_TrainVal}^{(k)}$), and the pseudo-labelled augmented data  ($D_{U}^{(k)}$) with evaluation restricted strictly to the held-out test partition ($D_{L\_test}^{(k)}$) for each fold $k$. Model selection was performed using the Validation Macro-F1 score, ensuring that the reported test metrics correspond to the epoch with the highest performance on the validation set.

\subsubsection{Results}

The proposed architecture is compared against two unimodal baselines: Feature Vector Only, Whisper Encoder Only and the Proposed hyrbid architecture combining the Whisper encoder embeddings with the feature vector.

\begin{table}[htbp]
\caption{ Mean Macro-F1 scores of the Hybrid Confidence Model}
\begin{center}
\begin{tabular}{l|c|c|c}
\toprule
\textbf{Confidence} & \textbf{FV Only} & \textbf{Whisper Only} & \textbf{Proposed} \\
\midrule
Low  & $0.666\pm 0.098$ & $0.714\pm 0.086$ & $\mathbf{0.744\pm 0.068}$ \\
Medium  & $0.532\pm 0.032$ & $0.656\pm 0.080$ & $\mathbf{0.672\pm 0.052}$\\
High  & $0.796\pm 0.032$ & $\mathbf{0.838\pm 0.041}$ & $0.836\pm 0.036$ \\
\midrule
\textbf{ Macro-F1} & $0.665\pm 0.041$ & $0.736\pm 0.049$ & $\mathbf{0.751\pm 0.041}$ \\
\bottomrule
\end{tabular}
\label{tab:ablationFV}
\end{center}
\end{table}

As shown in Table~\ref{tab:ablationFV}, the Whisper-only baseline outperforms the Feature Vector Only model ($0.736$ vs. $0.665$ Macro-F1). This performance gap is likely due to  the fundamental difference in capacity. The feature vector model is restricted to a limited set of 94 handcrafted descriptors whereas the Whisper encoder leverages deep, high-dimensional representations learned from 680k hours of diverse audio. Consequently, the Whisper embeddings likely capture a complex interplay of both phonetic and prosodic cues that simple feature engineering misses. Moreover, the feature vector model struggles most especially with the Medium Confidence class ($0.532$). This highlights the limitations of low-dimensional feature sets, which fail to capture the subtle dynamics required to distinguish confident from uncertain speech.

Despite the strong performance of the Whisper-Only baseline, the proposed hybrid architecture achieves the highest overall performance ($0.751$ Macro-F1), confirming that explicit acoustic supervision remains beneficial. Specifically, the hybrid model improves the minority class detection (low and medium confidence) compared to the Whisper baseline. This suggests that while Whisper implicitly captures prosody, it may miss subtle uncertainty markers (like jitter or hesitation) in favor of dominant linguistic patterns. The  feature vector resolves this by emphasising these specific signals that the Whisper-Only baseline missed.

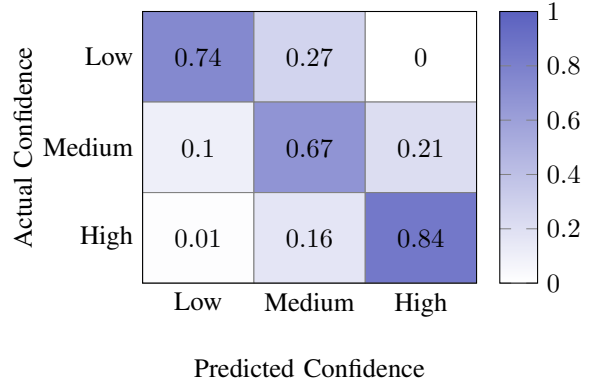
\begin{figure}
\begin{tikzpicture}
    \begin{axis}[
            width=6cm,
            colormap={bluewhite}{color=(white) rgb255=(90,96,191)},
            xlabel=Predicted Confidence,
            xlabel style={yshift=-10pt},
            ylabel=Actual Confidence,
            ylabel style={yshift=10pt},
            xticklabels={Low, Medium, High}, 
            xtick={0,1,2}, 
            xtick style={draw=none},
            yticklabels={Low, Medium, High}, 
            ytick={0,1,2}, 
            ytick style={draw=none},
            enlargelimits=false,
            colorbar,
            xticklabel style={rotate=0, yshift=0pt},
            nodes near coords={\pgfmathprintnumber[fixed, precision=2]{\pgfplotspointmeta}},
            nodes near coords style={yshift=-7pt},
            point meta min=0,
            point meta max=1,
        ]
        \addplot[
            matrix plot,
            mesh/cols=3, 
            point meta=explicit,
            draw=gray
        ] table [meta=C] {
            x y C
            0 0 0.74
            1 0 0.27
            2 0 0.00
            0 1 0.1
            1 1 0.67
            2 1 0.21
            0 2 0.01
            1 2 0.16
            2 2 0.84
        }; 
    \end{axis}
    
\end{tikzpicture}
\caption{Confusion Matrix for Hybrid Confidence Model}
\label{fig:conf_matrix}
\end{figure}

Fig.~\ref{fig:conf_matrix} presents the confusion matrix for the proposed hybrid model. The results demonstrate that the model is highly robust against catastrophic errors. Specifically, the misclassification of Low confidence samples as High confidence is non-existent ($0.00$) on the test set, and less than $1\%$ of High confidence samples are misclassified as Low confidence. Moreover, errors concentrate around the Medium confidence boundary, with 27\% of Low confidence samples predicted as Medium. This confirms the model has learned the ordinal nature of confidence, restricting misclassifications to neighbouring classes rather than making random predictions.

\begin{figure}

\centering
\includegraphics[width=7cm]{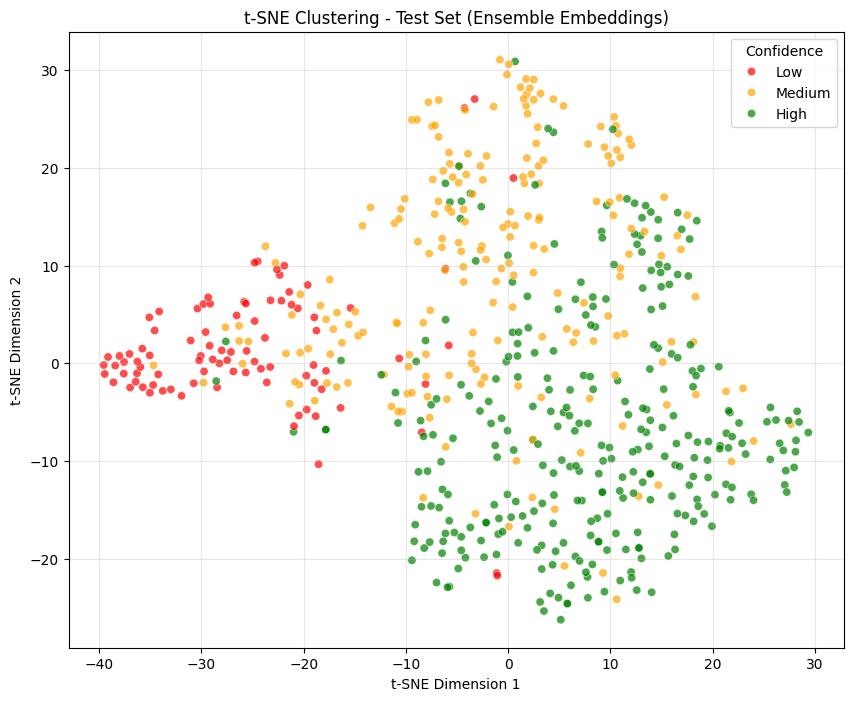}
\caption{t-SNE visualisation of test set embeddings using the ensemble model.} \label{fig:tsne}
\end{figure}

To further analyse the model's decision-making, a t-SNE projection of the embeddings for the ground truth data was plotted using the ensemble model, which also confirms that the model has learned a meaningful ordinal manifold. As seen in the t-SNE plot (Fig.~\ref{fig:tsne}), the Low Confidence (Red) and High Confidence (Green) clusters occupy distinct regions of the latent space with minimal overlap. The Medium Confidence class (Yellow) acts as a transition between low and high confidence, validating its role as an ambiguous boundary class where linguistic and paralinguistic cues naturally overlap.

\begin{figure*}[t]
\centering
\includegraphics[width=15cm]{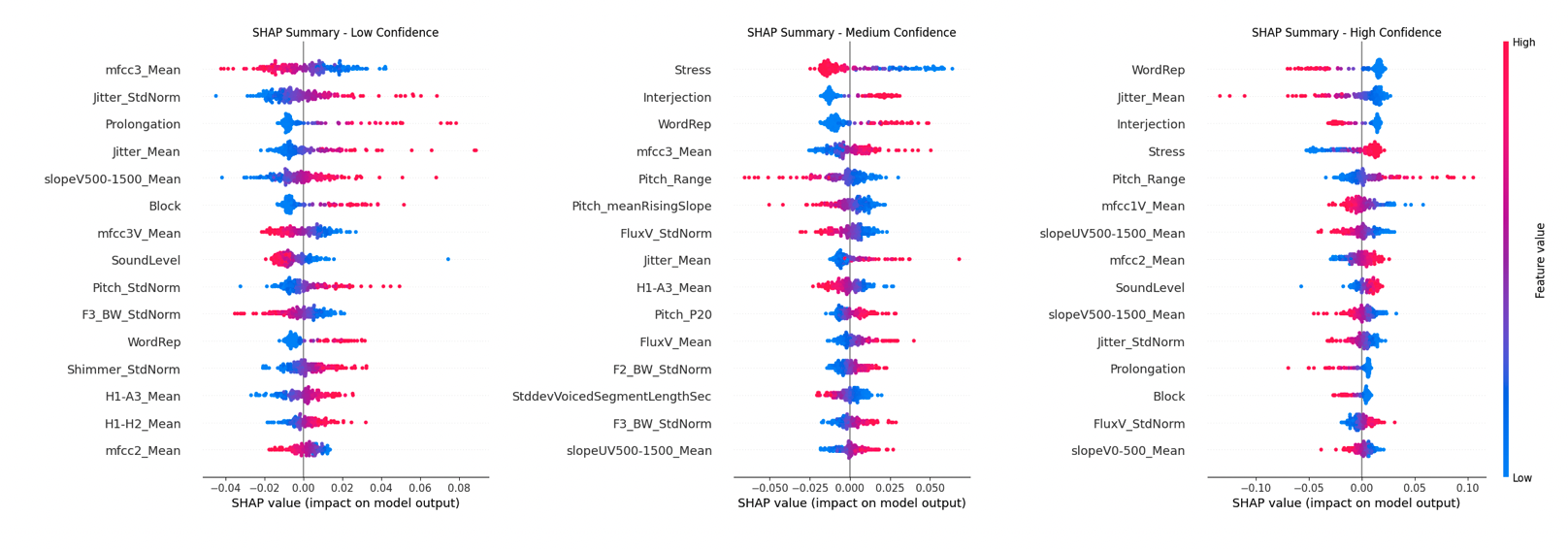}
\caption{SHAP Feature Importance Analysis for Low, Medium and High Confidence.} 
\label{fig:shap}
\end{figure*}

Moreover, to understand the specific acoustic drivers behind this clustering, the contribution of the  features in the feature vector were analysed using SHAP (SHapley Additive exPlanations) \cite{SHAP}. As shown in figure \ref{fig:shap}, the Low Confidence predictions are driven heavily by spectral characteristics and  vocal instability. Specifically, MFCC 3 Mean emerges as the top predictor, where low values strongly correlate with low confidence. MFCC 3 often relates to the quality of the vocal tract, where lower values result in a flatter shape often heard in mumbled speech. Jitter Standard Deviation and Jitter Mean, which are measures of pitch perturbation, are also amongst the most influential features, indicating that vocal tremor is a dominant signal for low confidence. Furthermore, disfluency types such as Prolongation and Block appear as strong positive predictors.

 For the Medium Confidence class, stress is the top feature for this class, with distinct clustering suggesting that specific stress patterns likely serve as key differentiators for this ambiguous middle ground, where lower stress values make the model more likely to classify speech as medium. Interjection and Word Repetition also appear as top positive predictors, suggesting that the model identifies Medium confidence through indicators of hesitation rather than vocal quality (e.g. MFCC Mean and Jitter) that characterise Low confidence.

 Finally, for the High Confidence class, the model primarily looks for the absence of disfluencies. The plot shows a strong negative relationship between disfluencies (eg. Word Repetition and Interjection) and high confidence where the presence of disfluencies makes the model less likely to predict high confidence. Consistent with the Low Confidence analysis, Jitter remains a top predictor, but exhibits an inverse relationship where the model strongly associates low jitter values with high confidence. Additionally, Pitch Range shows a positive correlation, where a wider pitch range, commonly associated with more confident speech \cite{Jiang2014}, contributes to a high confidence classification, and in contrast, a higher pitch range makes a sample less likely to be classified as Medium confidence. However, the correlation between Stress and High Confidence could be because stress labels are generated by mapping emotions to stress, and the resulting features capture acoustic traits that overlap with confident speech.

\begin{table*}[htbp]
\caption{Comparison of Different Model Architectures.}
\begin{center}
\begin{tabular}{l|c|c|c|c|c|c}
\toprule
\textbf{Architecture} & \textbf{Wav2Vec 2.0} & \textbf{HuBERT} & \textbf{WavLM} & \textbf{Whisper-Tiny} & \textbf{Wynn et al, 2025 \cite{WynnAIED2025}} & \textbf{Whisper-Base (Proposed)} \\
\midrule
Low  & $0.604\pm 0.129$ & $0.704\pm 0.068$ & $0.726\pm 0.083$ 
& $0.717\pm 0.089$ & $0.628\pm 0.185$ & $\mathbf{0.744\pm 0.068}$ \\
Medium  & $0.592\pm 0.088$ & $0.636\pm 0.030$ &$0.672\pm 0.025$ 
& $0.648\pm 0.057$ & $0.582\pm 0.100$ & $\mathbf{0.672\pm 0.052}$\\
High  & $0.788\pm 0.033$ & $0.806\pm 0.040$ & $0.814\pm 0.024$ 
& $0.818\pm 0.011$ & $0.732\pm 0.081$ & $\mathbf{0.836\pm 0.036}$ \\
\midrule
\textbf{Macro-F1} & $0.661\pm 0.062$ & $0.715\pm 0.032$ & $0.737\pm 0.019$ 
& $0.728\pm 0.033$ & $0.647\pm 0.053$ & $\mathbf{0.751\pm 0.041}$ \\
\bottomrule
\end{tabular}
\label{tab:ablationModel}
\end{center}
\end{table*}

\subsection{Impact of Training Strategies}
To assess the robustness of the semi-supervised pipeline, we conducted a series of studies regarding encoder adaptability and data filtering.

\subsubsection{Impact of Unfreezing Whisper Encoder Blocks}

Unfreezing all encoder blocks of Whisper-base resulted in a Macro-F1 of $0.742$ with class-wise scores of $0.739$ (Low), $0.658$ (Medium), and $0.830$ (High). Therefore, freezing the first 3 layers offered the optimal balance, achieving Macro-F1 of $0.751$. This suggests that the initial layers contain fundamental acoustic filters that are best left preserved, whilst the higher-level layers require adaptation to capture the semantic nuances of confidence.

\subsubsection{Impact of Semi-Supervised Data Strategy}
To validate the semi-supervised learning framework, the impact of pseudo-label quality versus quantity is examined.  The proposed uncertainty filtering approach ($\tau > 0.8$; $N\approx 1194$) is compared against two ablations: (1) training on ground truth data only ($N = 480$), and (2) using all generated pseudo-labels without filtering ($N = 11069$).

\begin{figure}[t]
\centering
\begin{tikzpicture}
    \begin{axis}[
        ybar,
        bar width=8pt,
        width=\linewidth,
        height=5cm,
        enlarge x limits=0.2,
        ylabel={F1-Score},
        symbolic x coords={Low, Medium, High, Macro-Avg},
        xtick=data,
        ymin=0.4, ymax=1.0, 
        ymajorgrids=true,
        grid style=dashed,
        legend style={at={(0.5,1.15)}, anchor=north, legend columns=-1, draw=none, fill=none},
        error bars/y dir=both,
        error bars/y explicit,
    ]

    \addplot[fill=gray!30] coordinates {
        (Low,0.69) +- (0,0.087)  
        (Medium,0.645) +- (0,0.067)
        (High,0.844) +- (0,0.031)
        (Macro-Avg,0.726) +- (0,0.056)
    };

    \addplot[fill=red!40] coordinates {
        (Low,0.662) +- (0,0.139)   
        (Medium,0.556) +- (0,0.079)
        (High,0.818) +- (0,0.04)
        (Macro-Avg,0.685) +- (0,0.076)
    };

    \addplot[fill=blue!50] coordinates {
        (Low,0.744) +- (0,0.068)  
        (Medium,0.672) +- (0,0.052)
        (High,0.836) +- (0,0.036)
        (Macro-Avg,0.751) +- (0,0.041)
    };

    \legend{GT Only, PL (No Filter), Proposed}
    \end{axis}
\end{tikzpicture}
\caption{Impact of Data Strategy. Comparing Ground-Truth (GT) Only against indiscriminate pseudo-labelling (No Filter) and the proposed uncertainty-aware filtering. Error bars represent standard deviation across 5 folds. }
\label{fig:ablation_chart}
\end{figure}
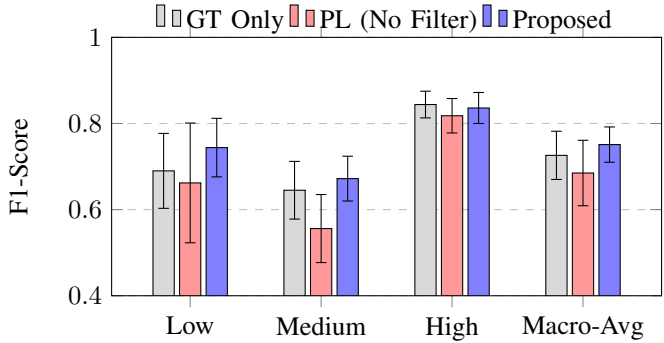

 As shown in Fig.~\ref{fig:ablation_chart}, training on the full, unfiltered pseudo-labelled (PL) dataset $D_U$ caused performance to fall to a Macro-F1 of 0.685. This could suggest that the noise and bias from the pseudo-labeller outweighs the benefits of additional labelled data, highlighting that high data quality is more important than simply increasing the amount of data available. On the other hand, relying exclusively on the ground truth (GT) data yielded a Macro-F1 of 0.726. This suggests that the model might have overfitted to the limited ground truth data, limiting its capacity to generalise to unseen speakers.

The proposed uncertainty filtering approach achieves the optimal F1 ($0.751$  by leveraging the scale of pseudo-labels while maintaining quality through strict filtering. This strategy proves effective especially for the minority classes (Low Confidence and Medium Confidence detection), demonstrating that high-quality, diverse data is essential for capturing the subtle acoustic variations of uncertainty that are underrepresented in the small ground truth set.

\subsubsection{Impact of Model Architecture}
To isolate the contribution of the underlying representation, the proposed Whisper-Base encoder was benchmarked against three state-of-the-art Self-Supervised Learning models (Wav2Vec 2.0 \cite{wav2vec2.0}, HuBERT \cite{HuBERT}, WavLM \cite{WavLM}) and a smaller architectural variant (Whisper-Tiny). Table~\ref{tab:ablationModel} summarises the results. Wav2Vec 2.0 yields the lowest performance ($0.661$ Macro-F1), but HuBERT and WavLM show notable gains, with WavLM achieving a strong score of $0.737$. Notably, WavLM matches the proposed model in the Medium confidence class ($0.672$).

However, the proposed Whisper-Base architecture achieves the highest overall stability and performance ($0.751$ Macro-F1). Specifically in the Low Confidence class, Whisper-Base outperforms WavLM. This superior generalisation on the minority class could suggest that the massive pre-training is beneficial. WavLM is trained on approximately 94k hours of unlabelled data \cite{WavLM}, whereas Whisper leverages 680k hours of weakly supervised audio, allowing the encoder to learn a far more robust feature space for difficult, ambiguous examples. Furthermore, increasing model capacity from Tiny ($0.728$) to Base ($0.751$) yields a consistent improvement across all metrics, confirming that both data scale and parameter count are necessary to capture the nuances of confidence.

Furthermore, the proposed architecture is compared against our preliminary work \cite{WynnAIED2025} which used a co-attention mechanism to fuse the acoustic feature vector with the Whisper embeddings prior to a downstream sequence model. However, as shown in Table~\ref{tab:ablationModel}, the framework proposed in this paper achieves a higher Macro-F1 score despite being architecturally simpler indicating that the previous co-attention mechanism unnecessarily increased model complexity and the proposed method  achieves superior performance with greater efficiency.

\section{Discussion}
\label{sec:discussion}

The analysis results in Section IV  suggest that the superior performance of Whisper-Base over purely acoustic self-supervised learning models such as WavLM could be due to the scale of pre-training. However, it could also be due to the nature of the training objective for these models. Unlike masked acoustic prediction, Whisper's ASR objective inherently aligns audio with linguistic semantics. This likely enables the model to learn implicit lexical information that correlates with confidence, which models trained on acoustic reconstruction will not see.

However, relying on these implicit linguistic cues creates a vulnerability when the spoken words contradict the speaker's tone. For example, a declarative statement could be misclassified as High Confidence by the baseline if it relies too heavily on the confident phrasing. The improved performance of the Hybrid model implies that the added acoustic features likely provide a corrective signal in these ambiguous instances.

To answer \textbf{RQ1}, we demonstrated that pseudo-labelling is an effective strategy for upscaling small labelled datasets. The proposed framework achieved a Macro-F1 of 0.751, outperforming the ground-truth baseline of 0.726. Moreover regarding \textbf{RQ2}, our ablation studies confirm that data quality supersedes quantity. We found that indiscriminate pseudo-labelling degraded performance compared to the baseline. This aligns with findings in \cite{arazo2020pseudo} regarding confirmation bias where without filtering, the model reinforces its own errors on noisy samples and the Uncertainty-Aware strategy ($\tau > 0.8$) demonstrates that data quality supersedes quantity. By restricting the training set to high-confidence samples, the model was more able to learn robust decision boundaries without overfitting to the small ground truth dataset ($N=600$).

Addressing \textbf{RQ3}, we established that fusing explicit paralinguistic features with Whisper embeddings yields a measurable performance gain over unimodal baselines. Our analysis shows that whilst Whisper provides a robust foundation, the explicit acoustic features act as a necessary corrective mechanism, allowing the model to distinguish between levels of confidence. For example, the SHAP plot (Fig. \ref{fig:shap}) suggests that the model actively uses Jitter and the presence of Word repetitions to distinguish Medium from High confidence.  This confirms that the feature vector acts as a corrective mechanism, preventing the semantic bias observed in the Whisper-only baseline.

Whilst the proposed architecture demonstrates strong performance in confidence estimation, several limitations exist. The primary constraint is the size of the ground truth dataset ($N=600$). Although we mitigated this scarcity through semi-supervised learning, uncertainty filtering and pseudo-labelling, the core validation remains bound to a relatively small set of annotations on English-speaking audio meaning that the model may not generalise to speakers outside this distribution. Future work must validate these findings on larger, more diverse corpora to ensure cross-demographic fairness.

Furthermore, our model processes short, isolated audio clips between 5 and 12 seconds. Therefore, the model lacks the  context needed to distinguish genuine confidence from more performative behaviours such as sarcasm. It also misses important cues from earlier in the dialogue and struggles with instances where confidence fluctuates  within a single clip. Moreover, without demographic context, the model may misinterpret cultural differences in pacing or hesitation as signs of a lack of confidence rather than differences in how confidence is portrayed and received.

Finally, in this study we only focus on audio but confidence is an inherently multimodal phenomenon. Research indicates that non-verbal visual cues such as eye contact, posture, and facial micro-expressions are often stronger predictors of confidence than voice alone \cite{Mori2019}. By relying exclusively on the audio modality, our system is blind to scenarios where the voice is steady but the body language signals doubt. Integrating visual features would likely resolve many of the ambiguities particularly observed in the Medium confidence class where acoustic cues are subtle or conflicting.

\section{Conclusion}

In conclusion, this paper proposed a robust semi-supervised framework for the automatic detection of speaker confidence, addressing the dual challenges of data scarcity and label subjectivity. By introducing an uncertainty-aware pseudo-labelling mechanism, we demonstrated that it is possible to leverage unlabelled data to train reliable classifiers, provided that the training data is filtered using only high-quality data.

The analysis identifies Whisper-Base as a superior foundation compared to acoustic-only baselines like WavLM, likely due to its massive pre-training scale and implicit semantic alignment. Furthermore, our hybrid approach uses the feature vector to correct Whisper’s understanding. By adding these specific details, the model can adjust its prediction, ensuring that subtle cues in the voice are not overlooked by Whisper. The ablation studies reveal that data quality supersedes quantity, and that strictly filtering pseudo-labels ($\tau > 0.8$) outperformed training on the full, noisy dataset. This confirms that for subjective tasks, training data quality is more important than indiscriminate large-scale augmentation.
.

Future work will focus on the three main limitations in Section V. First, to address demographic biases, we will validate the framework on multilingual datasets and investigate how confidence markers differ across cultural groups. Moreover, we aim to integrate visual modalities (e.g., facial expressions and eye contact) to resolve the ambiguity of the Medium confidence class, we aim to move beyond analysing short, isolated clips by incorporating temporal context to measure how a speaker's confidence evolves over time.

\section*{Acknowledgments}
The authors sincerely thank the seven annotators for their contribution to the dataset curation.



\bibliographystyle{IEEEtran}

\section{Biography Section}
\vspace{-1.2cm}
\begin{IEEEbiographynophoto}{Adam Wynn}
is a PhD student in the Department of Computer Science at Durham University. He is interested in AI in education, automatic feedback, adaptive learning and educational technology. 
\end{IEEEbiographynophoto}

\begin{IEEEbiographynophoto}{Jingyun Wang}
is an assistant professor in the Department of Computer Science at Durham University. Her research focuses on harnessing the power of AI and HCI to drive innovation in the fields of education and digital health. 
\end{IEEEbiographynophoto}

\vfill

\end{document}